\documentclass{svmult}

\usepackage{graphicx}    
\usepackage{multicol}    

\def\be{\begin{equation}}
\def\ee{\end{equation}}
\def\Re{\mathop{\rm Re}}		
\def\Tr{\mathop{\rm Tr}}		

\begin{document}

\title*{Monte Carlo Overrelaxation for $SU(N)$ Gauge Theories}
\titlerunning{$SU(N)$ overrelaxation}
\author{Philippe de Forcrand\inst{1,2}\and
Oliver Jahn\inst{1,3}}
\institute{Institute for Theoretical Physics, ETH, CH-8093 Z\"urich, Switzerland
\and
CERN, Physics Department, TH Unit, CH-1211 Geneva 23, Switzerland
\texttt{forcrand@phys.ethz.ch}
\and
Center for Theoretical Physics, MIT, Cambridge, MA 02139, USA
\texttt{jahn@mit.edu}}
%
%
\maketitle

\abstract{The standard approach to Monte Carlo simulations of $SU(N)$ Yang-Mills theories updates
successive $SU(2)$ subgroups of each $SU(N)$ link. We follow up on an old proposal of Creutz, to perform overrelaxation in the full $SU(N)$ group instead, and show that it is more efficient.}

\bigskip

The main bottleneck in Monte Carlo simulations of QCD is the inclusion of light dynamical fermions.
For this reason, algorithms for the simulation of Yang-Mills theories have received less attention.
The usual combination of Cabibbo-Marinari pseudo-heatbath \cite{Cabibbo-Marinari} and Brown-Woch
microcanonical overrelaxation \cite{Brown-Woch} of $SU(2)$ subgroups is considered satisfactory. However, the large-$N$
limit of QCD presents a different perspective. Fermionic contributions are suppressed as $1/N$,
so that studying the large-$N$ limit of Yang-Mills theories is interesting in itself. High precision is
necessary to isolate not only the $N\to \infty$ limit, but also the leading $1/N$ correction. Such
quantitative studies by several groups are underway \cite{SUN}. They show that dimensionless combinations
of the glueball masses, the deconfinement temperature $T_c$, and the string tension $\sigma$
approach their $N\to \infty$ limit rapidly, with rather small corrections $\sim 1/N^2$, even down to
$N=2$. The prospect of making ${\cal O}(1/N^2)\sim 10\%$, or even ${\cal O}(1/N)\sim 30\%$ accurate predictions for real-world QCD is tantalizing. Numerical simulations can guide theory
and help determine the $N=\infty$ ``master field''.
Already, an old string prediction $T_c/\sqrt{\sigma}(N=\infty) = \sqrt{\frac{3}{\pi (d-2)}}$, first dismissed by the author himself because it disagreed with Monte Carlo data at the time \cite{Olesen},  appears to be accurate to within 1\% or better. Proposals about the force between charges of $k$ units of $Z_N$ charge, so-called $k$-string tensions, can be confronted with numerical simulations, which may or may not give support to connections between QCD and supersymmetric theories \cite{Armoni}.
Efficient algorithms for $SU(N)$ Yang-Mills theories are highly desirable.

Here, we revive an old, abandoned proposal of Creutz \cite{Creutz}, to perform overrelaxation in the full $SU(N)$
group, and show its superiority over the traditional $SU(2)$ subgroup approach\footnote{Global updates, of Hybrid Monte Carlo type, are not considered here, because they were shown to be ${\cal O}(100)$ times less efficient than local ones for $SU(3)$ in \cite{Sharpe}.}
.
\section{State of the Art}
\label{sec:1}
We consider the problem of updating a link matrix $U \in SU(N)$, from an old value $U_{\rm old}$ to $U_{\rm new}$, according to the probability density
\be
\label{prob}
P(U) \propto \exp(\beta \frac{1}{N} \Re \Tr X^\dagger U) \quad .
\ee
$\frac{1}{N} \Re\Tr X^\dagger U$ is the ``local action''. The matrix $X$ represents the sum of the
``staples'', the neighboring links which form with $U$ a closed loop contributing to the action.
This is the situation for the Wilson plaquette action, or for improved actions (Symanzik, Iwasaki, ...)
containing a sum of loops all in the fundamental representation. Higher representations make the local action non-linear in $U$. This typically restricts the choice of algorithm to Metropolis, although the approach below can still be used to construct a Metropolis candidate (as e.g. in \cite{Necco}).
Thus, $X$ is a sum of $SU(N)$ matrices, i.e. a general $N\times N$ complex matrix.

Three types of local Monte Carlo algorithms have been proposed: \\
$\bullet$ {\bf Metropolis}: a random step $R$ in $SU(N)$ is proposed, then accepted or rejected.
Thus, from $U_{\rm old}$, a candidate $U_{\rm new} = R U_{\rm old}$ is constructed.
To preserve detailed balance, the Metropolis acceptance probability is
\be
\label{Met_acc}
P_{\rm acc} = \min(1,\exp(\beta \frac{1}{N} \Re\Tr X^\dagger (U_{\rm new} - U_{\rm old}))) \quad .
\ee
Acceptance decreases as the stepsize, measured by the deviation of $R$ from the identity,
increases. And an $N\times N$ matrix multiplication must be performed to construct $U_{\rm new}$,
which requires ${\cal O}(N^3)$ operations. This algorithm is simple but inefficient, because the
direction of the stepsize is random. By carefully choosing this direction, a much larger step can be
taken as we will see. \\
$\bullet$ {\bf Heatbath}: a new matrix $U_{\rm new}$ is generated directly from the probability density $P(U)$ Eq.(\ref{prob}).
This is a manifest improvement over Metropolis, since $U_{\rm new}$ is completely independent of $U_{\rm old}$. However, sampling $P(U)$ requires knowledge of the normalization on the right-hand side of Eq.(\ref{prob}). For $SU(2)$, the simple algorithm of \cite{Creutz-SU2} has been perfected for large $\beta$ \cite{KP-HF}. For $SU(3)$, a heatbath algorithm also exists \cite{Pietarinen}, although it can hardly be called practical. For $SU(N), N>2$, one performs instead a pseudo-heatbath \cite{Cabibbo-Marinari}. Namely, the matrix $U_{\rm old}$ is multiplied by an embedded $SU(2)$ matrix
$R = {\bf 1}_{N-2} \otimes R_{SU(2)}$, chosen by $SU(2)$ heatbath from the resulting probability
$\propto \exp(\beta \frac{1}{N} \Re\Tr (X^\dagger U_{\rm old}) R)$.
Note that computation of the 4 relevant matrix elements of $(X^\dagger U_{\rm old})$ requires
${\cal O}(N)$ work. To approach a real heatbath and decrease the correlation of $U_{\rm new} = U_{\rm old}  R$ with $U_{\rm old}$, a sequence of $SU(2)$ pseudo-heatbaths is usually performed,
where the $SU(2)$ subgroup sweeps the $\frac{N(N-1)}{2}$ natural choices of off-diagonal elements of
$\tilde{U}$. The resulting amount of work is then ${\cal O}(N^3)$, which remains constant relative to
the computation of $X$ as $N$ increases. \\
$\bullet$ {\bf Overrelaxation}. Adler introduced stochastic overrelaxation for multi-quadratic actions \cite{Adler}. The idea is to go beyond the minimum of the local action and multiply this step by $\omega \in [1,2]$, ``reflecting'' the link $U_{\rm old}$ with respect to the action minimum. This results in faster decorrelation, just like it produces faster convergence in linear
systems. In fact, as in the latter, infrared modes are accelerated at the expense of ultraviolet modes,
as explained in \cite{Heller-Neuberger}. The overrelaxation parameter $\omega$ can be tuned. Its optimal value approaches 2 as the dynamics becomes more critical. In this limit, the UV modes do not
evolve and the local action is conserved, making the algorithm microcanonical. In practice, it is simpler to fix $\omega$ to 2, and alternate overrelaxation with pseudo-heatbath in a tunable proportion (typically 1 HB for 4-10 OR, the latter number increasing with the correlation length). In $SU(2)$, this strategy
has been shown to decorrelate large Wilson loops much faster than a heatbath \cite{Decker}, with
in addition some slight reduction in the amount of work. It is now the adopted standard. For $SU(N)$,
one performs $\omega=2$ microcanonical overrelaxation steps in most or all $SU(2)$ subgroups, as
described in \cite{Brown-Woch}.

The $SU(2)$ subgroup overrelaxation of Brown and Woch is simple and elegant. Moreover, it
requires minimal changes to an existing pseudo-heatbath program. But it is not the only possibility.
Creutz \cite{Creutz} proposed a general overrelaxation in the $SU(N)$ group. And Patel \cite{Patel} implemented overrelaxation in $SU(3)$, whose efficiency was demonstrated in \cite{Sharpe}.
Here, we generalize Patel's method to $SU(N)$.

\begin{figure}[t]
\centering
\includegraphics[height=6cm]{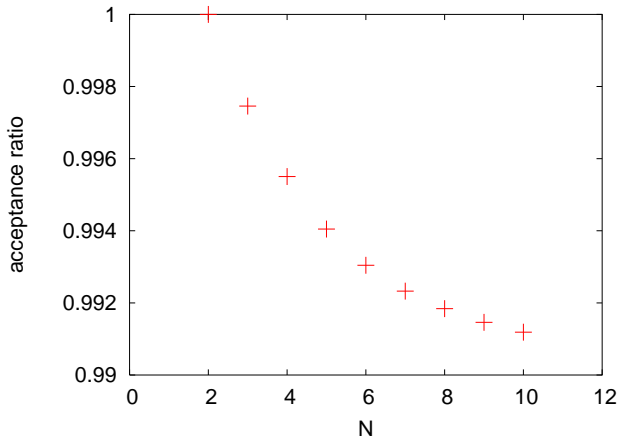}
\caption{$SU(N)$ overrelaxation acceptance versus $N$.}
\label{fig:1}
\end{figure}

\section{$SU(N)$ Overrelaxation}
\label{sec:2:pdf}

It may seem surprising at first that working with $SU(N)$ matrices can be as efficient as working on
$SU(2)$ subgroups. One must bear in mind that the calculation of the ``staple'' matrix $X$ requires
${\cal O}(N^3)$ operations, since it involves multiplying $N\times N$ matrices. The relative cost of
updating $U$ will remain bounded as $N$ increases, if it does not exceed ${\cal O}(N^3)$ operations.
An update of lesser complexity will use a negligible fraction of time for large $N$, and can be viewed
as a wasteful use of the staple matrix $X$. Therefore, it is a reasonable strategy to spend ${\cal O}(N^3)$ operations on the link update.  A comparison of efficiency should then be performed between
$(i)$ an update of all $\frac{N(N-1)}{2}$ $SU(2)$ subgroups, one after the other, following Cabibbo-Marinari and Brown-Woch; $(ii)$ a full $SU(N)$ update, described below, involving a polar decomposition of similar ${\cal O}(N^3)$ complexity. One may still worry that $(ii)$ is unwise because
the final acceptance of the proposed $U_{\rm new}$ will decrease very fast as $N$ increases.
Fig.~1 addresses this concern: the acceptance of our $SU(N)$ update scheme decreases in fact very slowly with $N$, and remains almost 1 for all practical $N$ values.

We now explain how to perform $SU(N)$ overrelaxation, along the lines of \cite{Creutz}.
The idea of overrelaxation is to go, in group space, in the direction which minimizes the action,
but to go beyond the minimum, to the mirror image of the starting point. If $\hat{X}$ is the
$SU(N)$ group element which minimizes the action, then the rotation from $U_{\rm old}$ to
$\hat{X}$ is $(\hat{X} U_{\rm old}^{-1})$. Overrelaxation consists of applying this rotation twice:
\be
\label{SUN_OR}
U_{\rm new} = (\hat{X} U_{\rm old}^{-1})^2 U_{\rm old} = \hat{X} U_{\rm old}^\dagger \hat{X}
\ee
$U_{\rm new}$ should then be accepted with the Metropolis probability Eq.(\ref{Met_acc}).
The transformation Eq.(\ref{SUN_OR}) from $U_{\rm old}$ to $U_{\rm new}$ is
an involution (it is equal to its inverse). From this property, detailed balance follows.

Note that this holds for {\em any} choice of $\hat{X}$ which is independent of $U_{\rm new/old}$,
resulting always in a valid update algorithm. Its efficiency, however, depends on making a clever
choice for $\hat{X}$. The simplest one is $\hat{X} = \bf 1 ~ \forall X$, but the acceptance is small.
Better alternatives, which we have tried, build $\hat{X}$ from the Gram-Schmidt orthogonalization
of $X$, or from its polar decomposition. We have also considered applying Gram-Schmidt or polar decomposition to $X^\dagger$ or to $X^*$. In all cases, a subtle issue is to make sure that $U_{\rm new}$ is indeed special unitary ($\det U_{\rm new} = 1$), which entails cancelling in $\hat{X}$ the phase usually present in $\det X$. The best choice for $\hat{X}$ balances work, Metropolis acceptance and effective stepsize.  Our numerical experiments have led us to the algorithm below, based on the polar decomposition of $X$, which comes very close to 
finding the $SU(N)$ matrix which minimizes the
local action. Note that Narayanan and Neuberger \cite{N-N} have converged independently to
almost the same method (they do not take Step 3 below).

\subsection{Algorithm}

\noindent
{\bf 1}. Perform the Singular Value Decomposition (SVD) of $X$: $X = U \Sigma V^\dagger$, where $U$ and $V~\in U(N)$, and $\Sigma$ is the diagonal matrix of singular values $\sigma_i $ ($\sigma_i = \sqrt{\lambda_i}$, where the $\lambda_i$'s are the eigenvalues of the non-negative Hermitian matrix $X^\dagger X$). It is simple to show that $W \equiv U V^\dagger$ is the $U(N)$ matrix which maximizes
$\Re\Tr X^\dagger W$. Unfortunately, $\det U V^\dagger \neq 1$.

\noindent
{\bf 2}. Compute $\det X \equiv \rho \exp(i \phi)$. The matrix $\hat{X}_{NN} = \exp(-i \frac{\phi}{N}) U V^\dagger$ is a suitable $SU(N)$ matrix, adopted by Narayanan and Neuberger \cite{N-N}.

\noindent
{\bf 3}. Find an approximate solution $\{\theta_i\}$ for the phases of the diagonal matrix $D = {\rm diag}(\exp(i\theta_1),..,\exp(i\theta_N)), \sum_N \theta_i = 0 ~ {\rm mod} ~ 2\pi$, to maximize
$\Re\Tr X^\dagger (\exp(-i \frac{\phi}{N}) U D V^\dagger)$. To find an approximate solution to this
non-linear problem, we assume that all phases $\theta_i$ are  small, and solve the linearized
problem.

\noindent
{\bf 4}. Accept the candidate $U_{\rm new} = \hat{X} U_{\rm old}^\dagger \hat{X}$, where $\hat{X} = \exp(-i \frac{\phi}{N}) U D V^\dagger$, with probability Eq.(\ref{Met_acc})
\footnote{This corresponds to an overrelaxation parameter $\omega=2$. It may be possible to
make the algorithm more efficient by tuning $\omega$, using the LHMC approach of \cite{LHMC}.}.

\subsection{Efficiency}

We set out to compare the efficiency of the algorithm above with that of the standard $SU(2)$ subgroup approach, as a function of $N$. Going up to $N=10$ forced us to consider a very small, $2^4$ system only. We chose a fixed 't Hooft coupling $g^2 N = 8/3$, so that $\beta = \frac{2N}{g^2} = \frac{3}{4} N^2$.
This choice gives Wilson loop values varying smoothly with $N$, as shown in Fig.~2, and is representative of current $SU(N)$ studies. The Metropolis acceptance, as shown in Fig.~1, remains
very close to 1. It increases with the 't Hooft coupling.

\begin{figure}[t]
\centering
\includegraphics[height=6cm]{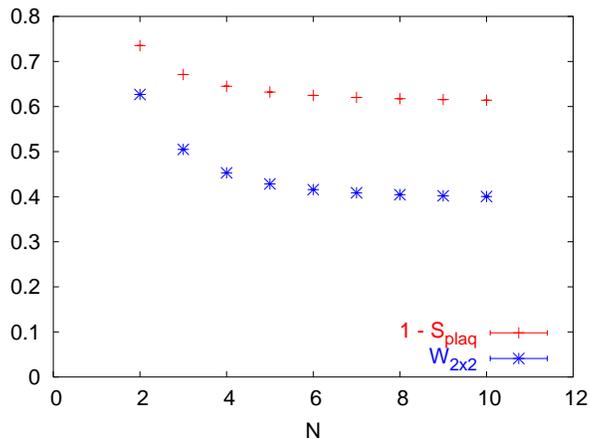}
\caption{Expectation values of 1x1 and 2x2 Wilson loops, versus $N$, for $\beta =\frac{3}{4} N^2$. }
\label{fig:2}
\end{figure}

\begin{figure}[t]
\centering
\includegraphics[height=6cm]{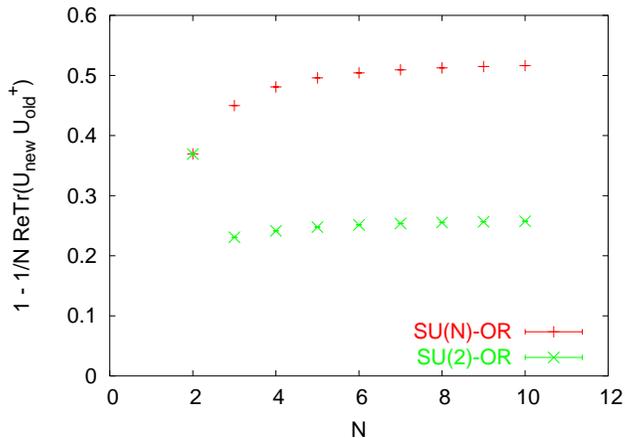}
\caption{Distance in group space after one link update, $\langle 1 - \frac{1}{N} \Re\Tr U_{\rm new}^\dagger U_{\rm old} \rangle$ vs $N$.}
\label{fig:3}
\end{figure}

A first measure of efficiency is given by the average stepsize, i.e. the link change under each update.
We measure this change simply by $\langle 1 - \frac{1}{N} \Re\Tr U_{\rm new}^\dagger U_{\rm old} \rangle$. The $SU(N)$ overrelaxation generates considerably larger steps than the $SU(2)$ subgroup
approach, as visible in Fig.~3. The real test, of course, is the decorrelation of large Wilson loops.
On our $2^4$ lattice, we cannot probe large distances. Polyakov loops (Fig.~4, left) show critical slowing down as $N$ increases, with a similar exponent $\sim 2.8$ using either update scheme.
The $SU(N)$ strategy gives a speedup ${\cal O}(3)$, more or less independent of $N$.
One observable, however, indicates a different dependence on $N$ for the two algorithms.
That is the asymmetry of the action, $\langle \sum_x \Re\Tr ({\rm Plaq^{timelike} - Plaq^{spacelike}}) \rangle$. Fig.~4, right, shows that the speedup provided by the $SU(N)$ overrelaxation grows like $\sim N^{0.55}$.
While this may be atypical, we never observed a slower decorrelation in the $SU(N)$ scheme for any observable.

\begin{figure}[t]
\centering
\includegraphics[height=4.0cm]{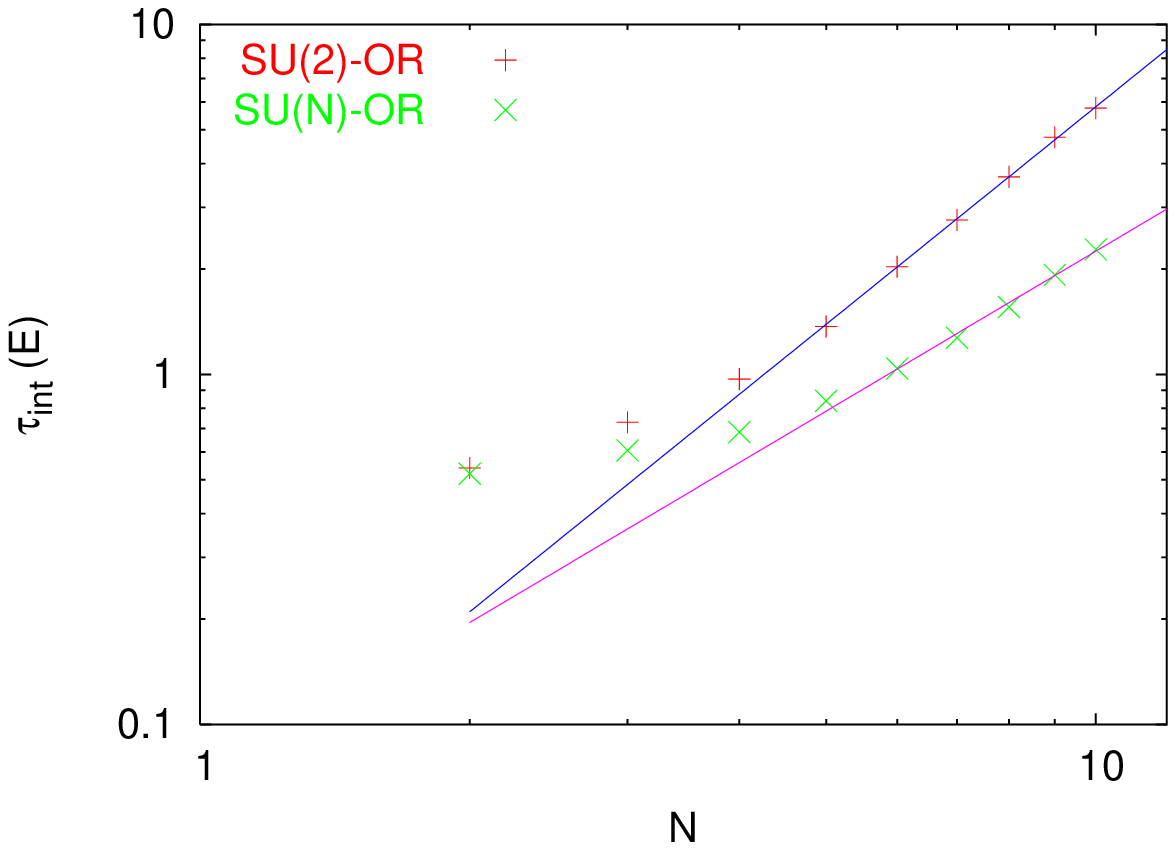}
\includegraphics[height=4.0cm]{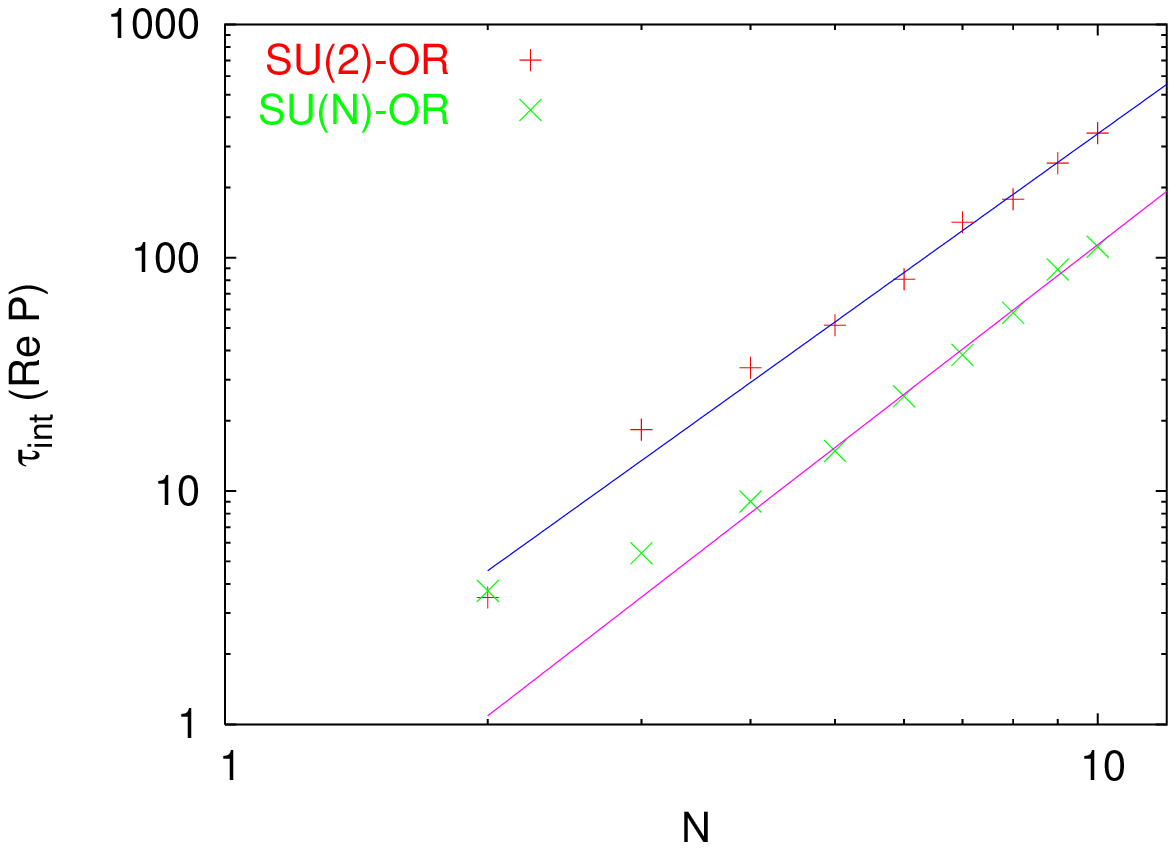}
\caption{Autocorrelation time of  Polyakov loop (left) and of $(S_{\rm timelike} - S_{\rm spacelike})$ (right) versus $N$, for the two algorithms.}
\label{fig:4}
\end{figure}

In conclusion, overrelaxation in the full $SU(N)$ group appears superior to the standard $SU(2)$ subgroup approach. The results of \cite{Sharpe} already indicated this for $SU(3)$. Our tests presented here suggest that the advantage grows with $N$, at least for some observables. For $SU(4)$ in (2+1) dimensions \cite{SU4}, the decorrelation of the Polyakov loop was $\sim 3$ times faster in CPU time, using $SU(N)$ overrelaxation, although our code implementation used simple calls to LAPACK routines, which are not optimized for operations on $4\times 4$ matrices. We definitely recommend $SU(N)$ overrelaxation for large-$N$ simulations.
\bigskip
\section*{Acknowledgements}
Ph. de F. thanks the organizers for a stimulating conference, and Brian Pendleton in particular for his extraordinary patience with the proceedings. The hospitality of the Kavli Institute for Theoretical Physics, where this paper was written, is gratefully acknowledged.



\begin{thebibliography}{10}

\bibitem{Cabibbo-Marinari}
Cabibbo, N. and Marinari, E.:
``A new method for updating SU(N) matrices in computer simulations of gauge
theories,''
Phys.\ Lett.\ B {\bf 119} (1982) 387.

\bibitem{Brown-Woch}
Brown, F.~R. and Woch, T.~J.:
``Overrelaxed heat bath and Metropolis algorithms for accelerating pure gauge
Monte Carlo calculations,''
Phys.\ Rev.\ Lett.\  {\bf 58} (1987) 2394.

\bibitem{SUN}
See, e.g.,
Lucini, B. and Teper, M.:
``Confining strings in SU(N) gauge theories,''
Phys.\ Rev.\ D {\bf 64} (2001) 105019
[arXiv:hep-lat/0107007];

Del Debbio, L., Panagopoulos, H., Rossi, P. and Vicari, E.:
``Spectrum of confining strings in SU(N) gauge theories,''
JHEP {\bf 0201} (2002) 009
[arXiv:hep-th/0111090].

\bibitem{Olesen}
Olesen, P.:
``Strings, tachyons and deconfinement,''
Phys.\ Lett.\ B {\bf 160} (1985) 408;
see also
Pisarski, R.~D. and Alvarez, O.:
``Strings at finite temperature and deconfinement,''
Phys.\ Rev.\ D {\bf 26} (1982) 3735.

\bibitem{Armoni}
Armoni, A., Shifman, M. and Veneziano, G.:
``From super-Yang-Mills theory to QCD: planar equivalence and its
implications,''
arXiv:hep-th/0403071.

\bibitem{Creutz}
Creutz, M.:
``Overrelaxation and Monte Carlo simulation,''
Phys.\ Rev.\ D {\bf 36} (1987) 515.

\bibitem{Sharpe}
Gupta, R., Kilcup, G.~W., Patel, A., Sharpe, S.~R. and de Forcrand, P.:
``Comparison of update algorithms for pure gauge SU(3),''
Mod.\ Phys.\ Lett.\ A {\bf 3} (1988) 1367.

\bibitem{Necco}
Hasenbusch, M. and Necco, S.:
``SU(3) lattice gauge theory with a mixed fundamental and adjoint plaquette
action: lattice artefacts,''
JHEP {\bf 0408} (2004) 005
[arXiv:hep-lat/0405012].

\bibitem{Creutz-SU2}
Creutz, M.:
``Monte Carlo study of quantized SU(2) gauge theory,''
Phys.\ Rev.\ D {\bf 21} (1980) 2308.

\bibitem{KP-HF}
Kennedy, A.~D. and Pendleton, B.~J.:
``Improved heat bath method for Monte Carlo calculations in lattice gauge
theories,''
Phys.\ Lett.\ B {\bf 156} (1985) 393;
Fabricius, K. and Haan, O.:
``Heat bath method for the twisted Eguchi-Kawai model,''
Phys.\ Lett.\ B {\bf 143} (1984) 459.

\bibitem{Pietarinen}
Pietarinen, E.:
``String tension in SU(3) lattice gauge theory,''
Nucl.\ Phys.\ B {\bf 190} (1981) 349.

\bibitem{Adler}
Adler, S.~L.:
``An overrelaxation method for the Monte Carlo evaluation of the partition
function for multiquadratic actions,''
Phys.\ Rev.\ D {\bf 23} (1981) 2901.

\bibitem{Heller-Neuberger}
Heller, U.~M. and Neuberger, H.:
``Overrelaxation and mode coupling in sigma models,''
Phys.\ Rev.\ D {\bf 39} (1989) 616.

\bibitem{Decker}
Decker, K.~M. and de Forcrand, P.:
``Pure SU(2) lattice gauge theory on 32**4 lattices,''
Nucl.\ Phys.\ Proc.\ Suppl.\  {\bf 17} (1990) 567.

\bibitem{Patel}
Patel, A., Ph.D. thesis, California Institute of Technology, 1984 (unpublished);
Gupta, R.,  Guralnik, G.,  Patel, A., Warnock, T. and Zemach, C.:
``Monte Carlo renormalization group for SU(3) lattice gauge theory,''
Phys.\ Rev.\ Lett.\  {\bf 53} (1984) 1721.

\bibitem{N-N}
Neuberger, H., private communication, and,
e.g.,
Kiskis, J., Narayanan, R. and Neuberger, H.:
``Does the crossover from perturbative to nonperturbative physics
become a phase transition at infinite N?,''
Phys.\ Lett.\ B {\bf 574} (2003) 65.
[arXiv:hep-lat/0308033].

\bibitem{LHMC}
Kennedy, A.~D. and Bitar, K.~M.:
``An exact local hybrid Monte Carlo algorithm for gauge theories,''
Nucl.\ Phys.\ Proc.\ Suppl.\  {\bf 34} (1994) 786
[arXiv:hep-lat/9311017];
Horvath, I. and Kennedy, A.~D.:
``The local Hybrid Monte Carlo algorithm for free field theory:
reexamining overrelaxation,''
Nucl.\ Phys.\ B {\bf 510} (1998) 367
[arXiv:hep-lat/9708024].

\bibitem{SU4}
de Forcrand, P. and Jahn, O.:
``Deconfinement transition in 2+1-dimensional SU(4) lattice gauge theory,''
Nucl.\ Phys.\ Proc.\ Suppl.\  {\bf 129} (2004) 709
[arXiv:hep-lat/0309153].

\end{thebibliography}
\end{document}